\documentclass[twocolumn,secnumarabic,amssymb, nobibnotes, aps, prd]{revtex4-1}
\usepackage{graphicx}
\usepackage{amsmath}
\usepackage{subfigure}
\usepackage{dcolumn}
\usepackage{bm}
\usepackage{float}
\usepackage{verbatim}
\addcontentsline{toc}{section}{Acknowledgement}
\begin{document}

\title{Long-distance continuous-variable quantum key distribution using separable Gaussian states}%
\author{Jian Zhou$^{1}$, Duan Huang$^{1,2}$, and Ying Guo$^{1}$}
\email[Corresponding author:]{guoyingcsu@sina.com}
\affiliation{$^1$School of Information Science and Engineering,
Central South University, Changsha {\rm 410083}, China \\
$^2$State Key Laboratory of Advanced Optical Communication Systems and Networks,
Department of Electronic Engineering, Shanghai Jiao Tong University, Shanghai {\rm 200240}, PR China.}
\date{\today}%
\begin{abstract}
Continuous-variable quantum key distribution (CVQKD) is considered to be an alternative to classical cryptography for secure communication.
However, its transmission distance is restricted to metropolitan areas, given that it is affected by the channel excess noise and losses.
In this paper, we present a scheme for implementing long-distance CVQKD using separable Gaussian states.
This tunable QKD protocol requires separable Gaussian states, which are squeezed and displaced, along with the assistance of classical communication and available linear optics components.
This protocol originates from the entanglement of one mode and the auxiliary mode used for distribution,
which is first destroyed by local correlated noises and restored subsequently by the interference of the auxiliary mode with the second distant separable correlated mode.
The displacement matrix is organized by two six-dimensional vectors and is finally fixed by the separability of the tripartite system.
The separability between the ancilla and Alice and Bob's system mitigates the enemy's eavesdropping, leading to
tolerating higher excess noise and achieving longer transmission distance.
\end{abstract}
\maketitle

\section{Introduction}
Quantum key distribution (QKD)~\cite{bennett1984quantum,PhysRevLett.67.661} enables two distant parties, conventionally called Alice and Bob, who have access to
an authenticated classical channel, to share secret keys in the presence of eavesdropper, Eve. The unconditional security of an ideal QKD protocol has been established
even if it is exposed to an adversary, who possesses unlimited computing power and technological capabilities~\cite{Mayers2001Unconditional,PhysRevLett.114.070501,Leverrier2017Security,RevModPhys.81.1301}.
Normally, QKD is divided into two kinds: discrete-variable (DV) QKD~\cite{PhysRevLett.67.661,PhysRevA.69.012309}, which relies on photon counting techniques,
and continuous-variable (CV) QKD~\cite{PhysRevLett.88.057902,RevModPhys.84.621,PhysRevA.96.022320,PhysRevA.95.042326}, which relies on coherent detection.
Equipped with the decoy state technique~\cite{PhysRevLett.94.230504}, DVQKD
can realize hundreds of kilometers of communication~\cite{PhysRevLett.117.190501}. With the help of a satellite, the transmission distance of QKD has been extended to $1200$ kilometers~\cite{Liao2017Satellite}.
Another branch of QKD, CVQKD, which has stable, reliable light resources and high detection efficiency, is more compatible with classical optical communications when compared to DVQKD~\cite{RevModPhys.84.621}.
However, despite all the advantages, CVQKD cannot yet replace DVQKD since its transmission distance is too short~\cite{Pirandola2015High,PhysRevA.93.022325}.
One reason for the short distance is the presence of the eavesdropper, Eve, who can perturb the quantum system using the most general strategies allowed by quantum mechanics.
Another one is that CVQKD schemes require a far more complicated error correction procedure, which further restricts the secure transmission distance.

Einstein associated entanglement with spooky action-at-a-distance~\cite{PhysRev.47.777}, which is different from the current view in quantum information theory that
regards entanglement as a physical resource. Entanglement~\cite{RevModPhys.81.865} has been widely applied to QKD~\cite{Epping2017Multi}, quantum dense coding~\cite{PhysRevA.92.052330}, quantum teleportation~\cite{Ren2017Ground}, entanglement swapping~\cite{PhysRevLett.119.170502} and beating classical communication complexity bounds~\cite{PhysRevA.72.050305}.
For example, global quantum operations can be implemented in quantum teleportation utilizing entanglement and classical communication. Great effort has been devoted to distributing
and manipulating entanglement among separated parties. In addition, a scheme of entangling two distant parties based on communication via a quantum channel and
local operations and classical communication (LOCC) was proposed~\cite{PhysRevLett.91.037902}.
Entanglement between distant parties can be created by sending a mediating particle between them via a quantum channel: swap the first particle with the ancilla, send it through the channel
and entangle it with the second particle. Besides the qubit protocol, distributing CV entanglement by separable Gaussian states has also been suggested~\cite{PhysRevA.77.050302,PhysRevA.80.032310}.
Two separable modes $A$ and $B$ may be entangled after interacting with the auxiliary mode $C$. Unfortunately, pure quantum states cannot achieve this target.
Moreover, Alice and Bob usually apply squeezing and displacement operations on these modes to enhance the practical quantum information processing. Recently, the aforementioned operations have been verified in experiment~\cite{PhysRevLett.111.230506,PhysRevLett.111.230504}.

To lengthen the transmission distance of the CVQKD system, we develop an improved protocol which transmits a separable ancilla without sending the secret information directly as usual.
It may entangle mode $A$, in Alice's laboratory, with separable mode $B$, in Bob's distant laboratory, by sending an ancillary mode $C$ which is separable from the subsystem $(AB)$~\cite{PhysRevLett.91.037902}.
Normally, the quantum transmission channel is assumed to be under Eve's control in QKD. We exemplify the entanglement between Alice's and Bob's modes
and the separability between the ancilla and the kept particle by calculating the lowest eigenvalue.
In previous fully Gaussian protocols, Eve's system $E$ purifies $AB$, so that, $S(E)=S(AB)$. Fortunately, in this scheme, the transmitted particle $C$ that may be attacked by
Eve is separable from $AB$. The eavesdropper cannot get access to Alice's and Bob's laboratories as well as the information transmitted in the classical channel.
In this case, it is impossible for the eavesdropper to recover the process of the protocol and hence she cannot extract any information.
In such a scenario, the proposed scheme reduces the information leaked to the eavesdropper, thus enables longer transmission distance.

This paper is organized as follows: In Sec.~\ref{sec2}, we review the distribution of entanglement with separable states.
In Sec.~\ref{sec3}, we present the details of CVQKD scheme with separable states. Sec.~\ref{sec4} shows the performance of
the proposed CVQKD scheme under general eavesdropping. Finally, we conclude this paper in Sec.~\ref{sec5}.

\section{Entanglement distribution with separable states}
\label{sec2}

\begin{figure}[htbp]
  \centering
  \includegraphics[height=45mm]{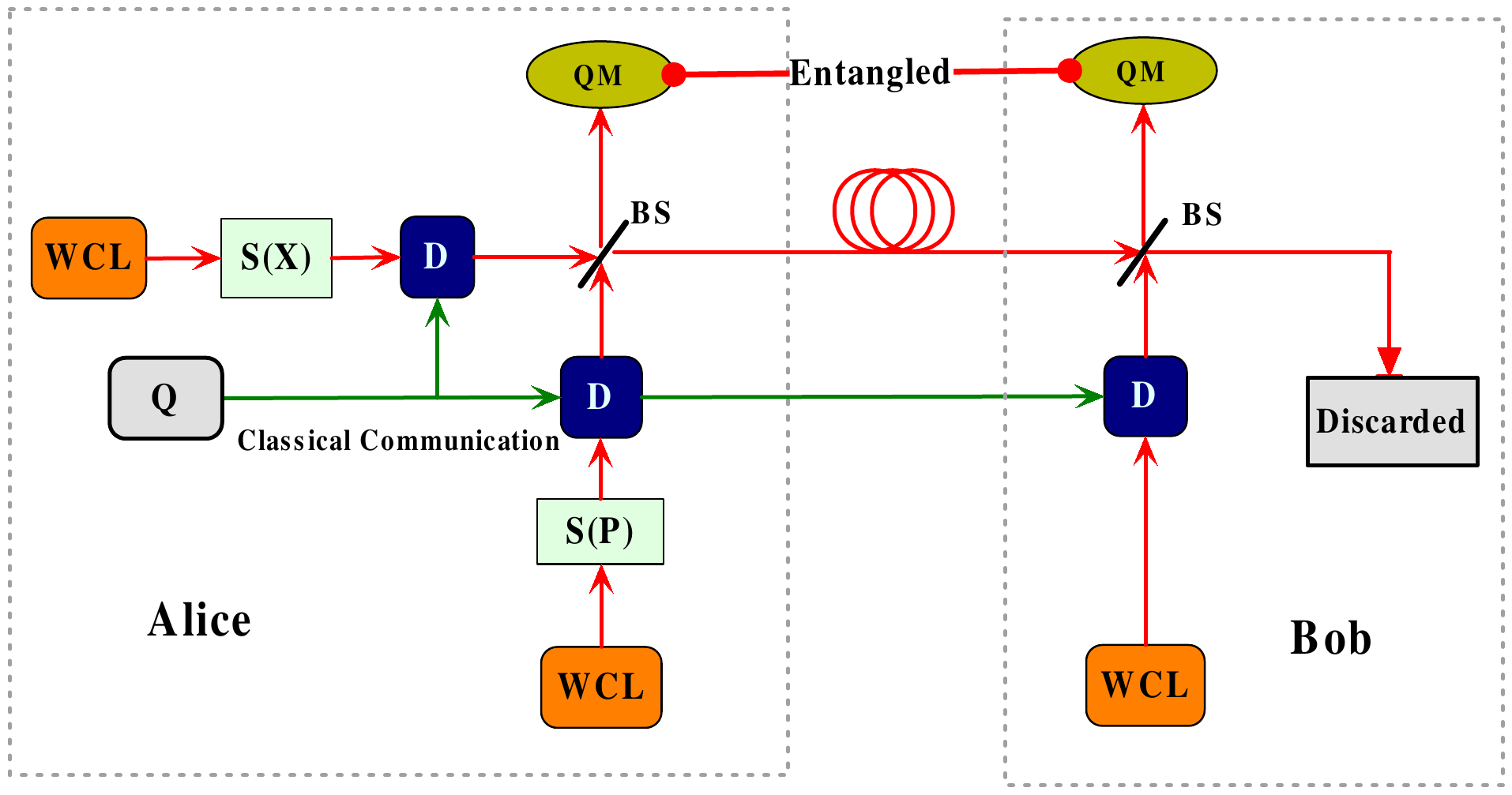}
\caption{(Color online) Alice's particle and Bob's particle interact with a mediating particle $C$ continuously.
 Alice and Bob get entangled while leaving $C$ separable from the system $AB$. WCL denotes weak coherent laser, and $S(X)$, $S(P)$ are compression operations on along position and momentum directions.
 $D$ is a local displacement distributed according to the Gaussian distribution with correlation matrix $Q$.}
  \label{figure1}
\end{figure}

Distributing entanglement with separable states is a breakthrough in the theory of quantum entanglement. It has been shown that separable Gaussian states
can be used for implementing entanglement distribution~\cite{PhysRevA.77.050302,PhysRevA.80.032310}. As shown in Fig.~\ref{figure1}, this process can be
accomplished by communication via a quantum channel and LOCC.

At the start of the original entanglement distribution protocol, Alice prepares systems $A$ and $C$ in a Gaussian state while Bob prepares system $B$ in a Gaussian state.
The three quantum systems are fully separable at this stage. Alice squeezes her two systems: one along the position quadrature and the other along the momentum quadrature.
In order to keep the ancilla separable from system $AB$, a displacement operation is applied to each of the three systems.
Note that the displacement is dependent on the squeezing parameters $r_1$ and $r_2$. Alice sends her two systems into a beam splitter.
The beam splitter operation on modes $A$ and $C$ results in a state separable with respect to two bipartitions: $B-AC$ and $C-AB$.
One of the outputs is stored in Alice's quantum memory (QM). The other is sent to Bob via a quantum channel. Bob also applies a
beam splitter operation on modes $B$ and $C$. Mixing of modes $B$ and $C$ on a balanced beam splitter finally entangles $A$ and $B$ while $C$ still remains separable
from $AB$.

In what follows, we recall how a displacement operation may make the transmitted ancilla $C$ separable from $AB$~\cite{PhysRevA.77.050302}. Before the displacement
operation, modes $A$ and $C$ are in a two-mode squeezed vacuum state and mode $B$ is in a vacuum state. The output of the first beam splitter is a two-mode squeezed vacuum state with
the following covariance matrix (CM):
\begin{align}
\gamma_{AC}=\left[
\begin{matrix}
\cosh{(2\tau)}I_2 & \sinh{(2\tau)}\sigma_z  \\
\sinh{(2\tau)}\sigma_z  &  \cosh{(2\tau)}I_2 \\
\end{matrix}
\right],
\end{align}
where $\tau\ \geq0$ is the squeezing parameter. Modes $A$ and $C$ are entangled when the lower symplectic eigenvalue $\nu_{\mathrm{min}}$ of the partial transpose of CM $\gamma_{AC}$
is less than one~\cite{PhysRevA.77.050302}. The CM of the three-mode system $ABC$ is given by
\begin{align}
\gamma_{ABC}=\left[
\begin{matrix}
\cosh{(2\tau)}I_2 & 0 & \sinh{(2\tau)}\sigma_z  \\
0 & I_2 & 0 \\
\sinh{(2\tau)}\sigma_z & 0 &  \cosh{(2\tau)}I_2 \\
\end{matrix}
\right].
\end{align}
We add an excess non-negative matrix $P$ to $\gamma_{ABC}$
\begin{align}
\gamma^1_{ABC}=\gamma_{ABC}+xP,
\label{3}
\end{align}
to entangle mode $A$ and modes $BC$, while leaving the other two bipartitions separable.
We follow the method for the construction of three-mode entangled Gaussian states in~\cite{PhysRevA.64.052303} to build matrix $P$.
The entanglement between modes $A$ and $C$ can be destroyed by adding a positive multiple of sum of the projectors onto the subspace
spanned by two six-dimensional vectors~\cite{PhysRevA.64.052303,PhysRevA.77.050302}. The negative eigenvalue of the CM is $\lambda=-(1-e^{-2\tau})$ with its eigenvector
$p_\lambda=p_1+ip_2$ for $p_1=(0,1,0,1)^T$ and $p_2=(1,0,-1,0)^T$. We extend $p_1$ and $p_2$ to the six-dimensional vectors
$q_1=(0,1,0,-2,0,1)^T$ and $q_2=(1,0,2,0,-1,0)^T$ with the displacement matrix $P=q_1q^T_1+q_2q^T_2$. In order to smear the entanglement
between modes $A$ and $C$, we add a sufficiently large, nonnegative multiple $xP$ to the CM as shown in Eq.~(\ref{3}) and obtain
\begin{align}
\gamma^1_{ABC}=\left[
\begin{matrix}
aI_2 & 2x\sigma_z & b\sigma_z  \\
2x\sigma_z & (1+4x)I_2 & -2xI_2 \\
b\sigma_z & -2xI_2 & aI_2 \\
\end{matrix}
\right].\label{eq2}
\end{align}
where $a=\cosh(2t)+x$ and $b=\sinh(2t)-x$.
Then the lowest symplectic eigenvalue of matrix $(\gamma^1_{ABC})^{(T_C)}$ can be derived as ~\cite{PhysRevA.65.032314},
\begin{align}
\nu_{\mathrm{min}}=\frac{\sqrt{(1+6x+e^{-2\tau})^2-32x^2}-(1+2x-e^{-2\tau})}{2}.
\end{align}
The separable bound of $C$ and $AB$ is $\frac{e^{2\tau}-1}{2}$, where the parameter $x$ should be equal or greater than this value.
On the other hand, the lowest eigenvalue of matrix $(\gamma^1_{ABC})^{(T_A)}$ can be calculated as
\begin{align}
\label{5}
\kappa_\mathrm{min}=\frac{1+6x+e^{-2\tau}-\sqrt{(1+2x-e^{2\tau})^2+32x^2}}{2}.
\end{align}
\begin{figure}
\centering
\subfigure[]{
\begin{minipage}[b]{0.48\textwidth}
\includegraphics[width=1\textwidth]{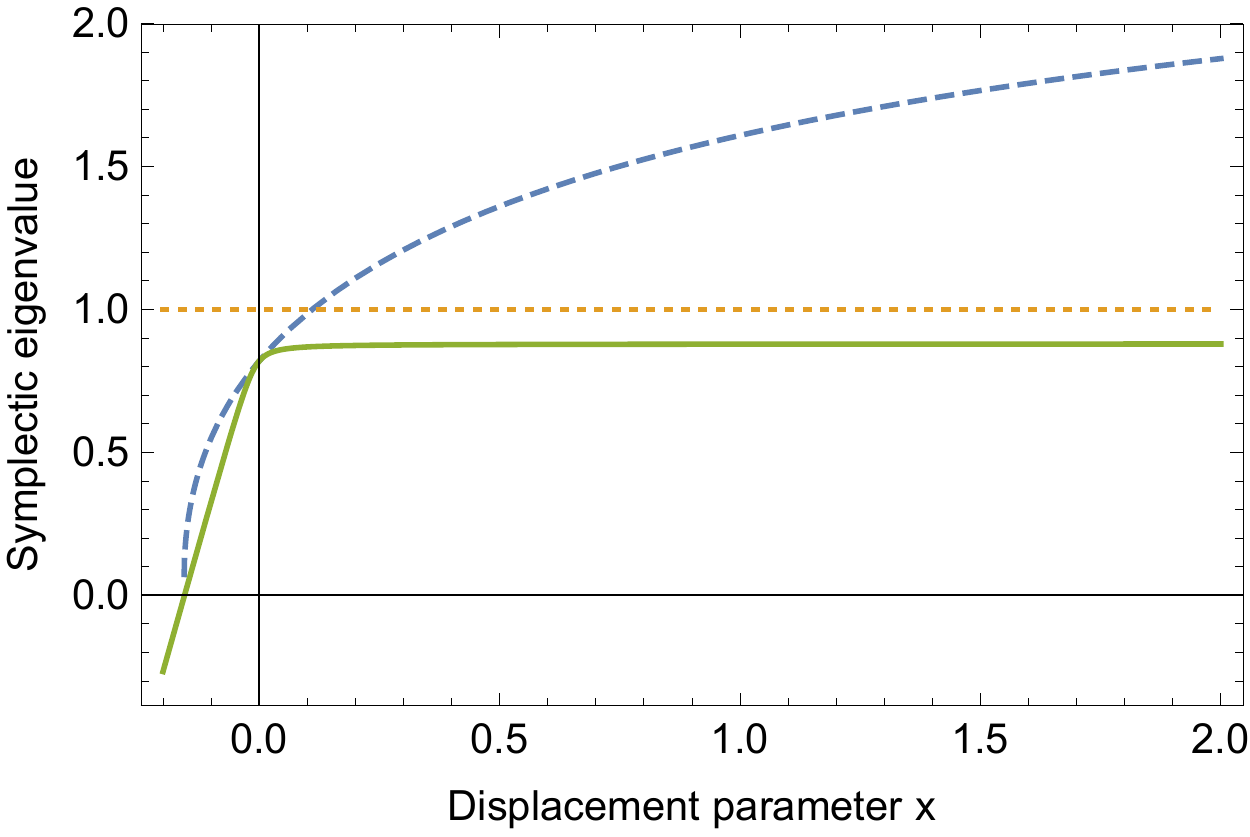}
\end{minipage}
}
\subfigure[]{
\begin{minipage}[b]{0.48\textwidth}
\includegraphics[width=1\textwidth]{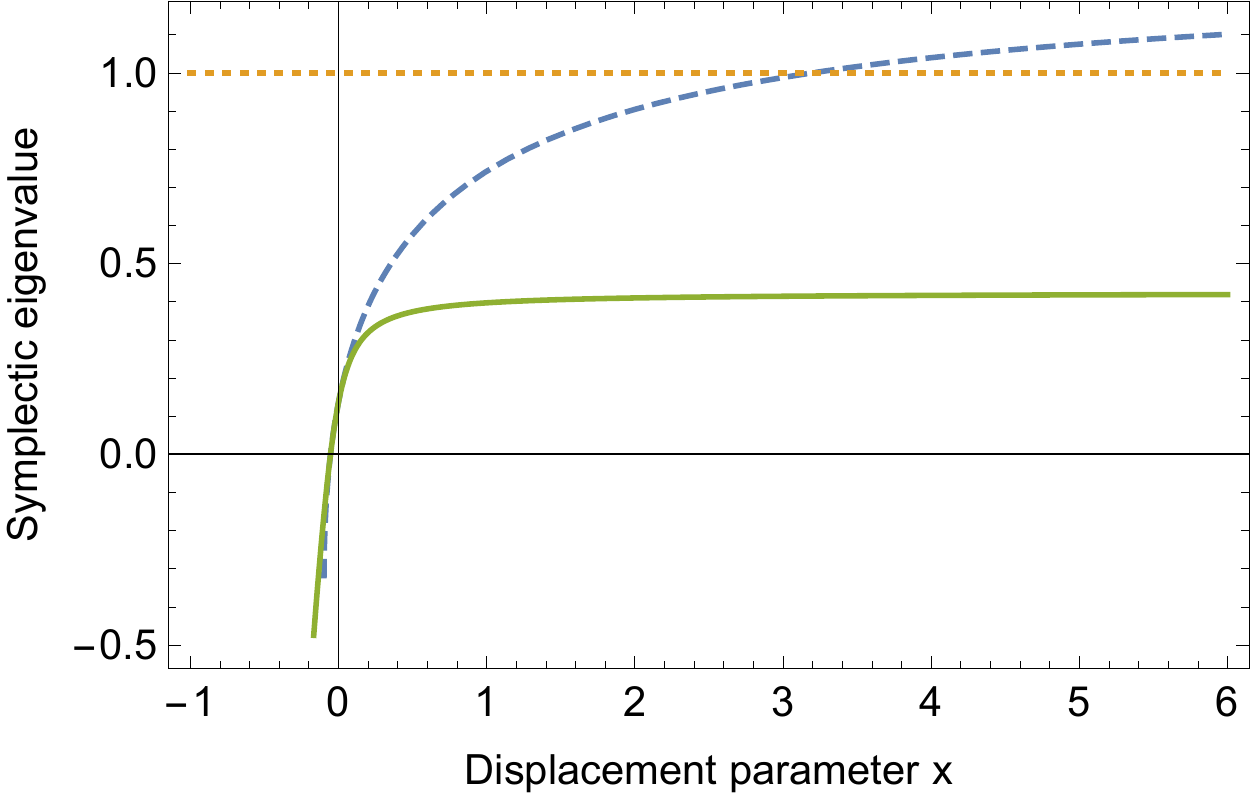}
\end{minipage}
}
 \caption{The eigenvalues, $\nu_{\mathrm{min}}$ and $\kappa_{\mathrm{min}}$, as a function displacement parameter $x$, for different compression parameters
  $\tau$, correspond to the dashed and full lines. The compression parameter $\tau=0.1$ in (a) and $\tau=1$ in (b). The dotted lines denote the boundary of separability.} \label{figure6}
\end{figure}
Taking $x\geq0$ and $\tau\geq0$ in Eq.~\ref{5}, the lowest eigenvalue is less than one, which verifies that there is entanglement between $A$ and $BC$.
Fig.~\ref{figure6} shows the lowest symplectic eigenvalue of matrix $(\gamma^1_{ABC})^{(T_C)}$ and $(\gamma^1_{ABC})^{(T_A)}$.
To satisfy the separability of $C-AB$, the lowest symplectic eigenvalue corresponding to the dashed line should be greater than one.
Similarly, the lowest symplectic eigenvalue corresponding to the full line ought to be less than one to ensure the entanglement between $A$ and $BC$.
Finally, after applying reverse operation of the beam splitter on $\gamma^1_{ABC}$, the covariance matrix of the random displacement distributed according to Gaussian
distribution is fixed. The beam splitter transforms the CM in~(\ref{eq2}) to CM $\gamma^{2}_{ABC}$ that is as follow:
\begin{align}
\gamma^2_{ABC}=\left[
\begin{matrix}
aI_2 & \frac{2x+b}{\sqrt{2}}\sigma_z & \frac{2x-b}{\sqrt{2}}\sigma_z  \\
\frac{2x+b}{\sqrt{2}}\sigma_z & \frac{1+a}{2}I_2 & \frac{1+4x-a}{2}I_2 \\
\frac{2x-b}{\sqrt{}2}\sigma_z & \frac{1+4x-a}{2}I_2 & \frac{1+8x+a}{2}I_2 \\
\end{matrix}
\right].\label{eq3}
\end{align}
The symplectic eigenvalue of CM $\gamma_{AB}$ can be calculated as $\nu=0.3968$ for $e^{2\tau}=10$, and the entanglement can be obtained as $E_\mathcal{N}=-\log_2{\nu}\approx1.33$ ebits.

According to the entanglement distribution with separable Gaussian states, we find that
the entanglement is firstly destroyed by displacement operations, which makes the auxiliary mode separable from sender's mode. After that, the auxiliary mode
is sent to Bob who partially restores the entanglement by mixing it with his suitably classically correlated mode, leading to the entanglement enhancement. Using this elegant characteristics, we propose
an improved CVQKD scheme to lengthen the maximum transmission distance with separable Gaussian states.

\section{Continuous variable quantum key distribution with separable Gaussian states}
\label{sec3}
This section is divided into three parts: the first part gives the CVQKD protocol using separable Gaussian states, the second part analyses the security of normal CVQKD protocol,
while the third subsection states the merit of the protocol based on separable Gaussian states.

\subsection*{A. Design of the CVQKD protocol using separable Gaussian states}
\begin{figure*}[!htp]
  \centering
  \includegraphics[height=60mm]{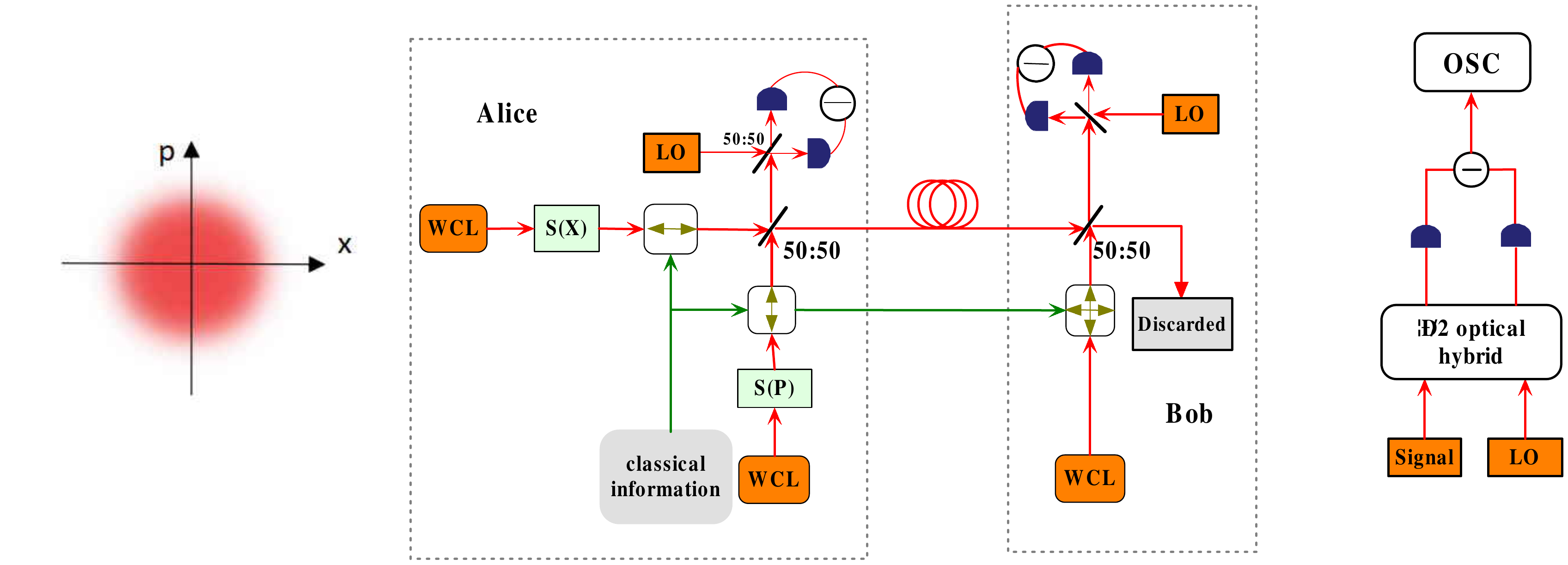}
  \caption{(Color online) Scheme of CVQKD by sending separable Gaussian states. Alice and Bob apply displacement operation on their
  state at the stage of preparation. The displacement ensures the separability between $C$ and $AB$. These modes emerge randomly in phase space obey Gaussian
  distribution as shown in the left part. The right part gives the detection scheme. WCL denotes weak coherent laser, and $S(X)$, $S(P)$ are compression operations along momentum and position directions.
  Double-headed arrow is local displacement distributed according to the correlation matrix.}
  \label{figure2}
\end{figure*}

Two normal parties, Alice and Bob aim to share secret key. For the sake of simplifying the process, we add the displacement operation
in the form of matrix while the practical displacement is not complex. The prepare and measure description of the CVQKD based on entanglement distribution protocol
using Gaussian states is shown in Fig.~\ref{figure2} and is described as follows.
\begin{itemize}
\item  Alice prepares two squeezed vacuum states which are position-squeezed and momentum-squeezed vacuum states, respectively. Displacement operations are added on these squeezed states.
The output of the first beam splitter is a two-mode squeezed vacuum state if we ignore the displacement operation.
\item  Alice detects one of the outputs with homodyne detection and sends another one to Bob via a quantum channel.
\item  After receiving Alice's mode, Bob interferes his vacuum state with the received state at a balanced beam splitter.
\item  Bob heterodynes one of the beam splitter's outputs with the self-referenced strategy, whereas another one is discarded directly.
\end{itemize}
In Alice's laboratory, she prepares two states, one position-squeezed vacuum state and one momentum-squeezed vacuum state given by
\begin{align}
\gamma_A=\left[
\begin{matrix}
e^{2\tau} & 0  \\
0 &  e^{-2\tau}
\end{matrix}
\right], &&\gamma_C= \left[\begin{matrix}
e^{-2\tau} &0\\
0 & e^{2\tau}
\end{matrix}\right].
\end{align}
The CM of the beam splitter's output can be expressed as
\begin{align}
\gamma_{AC}=\left[
\begin{matrix}
VI_2  & \sqrt{V^2-1}\sigma_z   \\
\sqrt{V^2-1}\sigma_z & VI_2 \\
\end{matrix}
\right],
\end{align}
with $V=\frac{e^{2\tau}+e^{-2\tau}}{2}$, $\sigma_Z=\left[\begin{smallmatrix}1&0 \\ 0&-1\\ \end{smallmatrix}\right]$ and $I_2=\left[\begin{smallmatrix}1&0 \\ 0&1\\ \end{smallmatrix}\right]$.
The CM of $ABC$ before transmission without displacement is
\begin{align}
\gamma_1=\left[
\begin{matrix}
VI_2  & 0 & \sqrt{V^2-1}\sigma_z \\
0 & I_2 & 0 \\
\sqrt{V^2-1}\sigma_z & 0 & VI_2 \\
\end{matrix}
\right].
\end{align}
Taking the displacement into consideration, the corresponding CM becomes
\begin{align}
\gamma_2=\left[
\begin{matrix}
aI_2  &  b\sigma_z & 2x\sigma_z \\
b\sigma_z & aI_2 & -2xI_2 \\
2x\sigma_z & -2xI_2 & (1+4x)I_2 \\
\end{matrix}
\right],
\end{align}
with $a=V+x$ and $b=\sqrt{V^2-1}-x$.
The linear channel can be equivalent to a beam splitter with transmittance $\eta$, the function of transmission distance $\eta=10^{-\frac{L}{50}}$.
The equivalent CM of the channel is
\begin{align}
B_\eta=\left[
\begin{matrix}
I_2 & 0 & 0 & 0 \\
0 & I_2 & 0 & 0 \\
0 & 0 & \sqrt{\eta}I_2  & \sqrt{1-\eta}I_2  \\
0 & 0 & -\sqrt{1-\eta}I_2 & \sqrt{\eta}I_2 \\
\end{matrix}
\right].
\end{align}
After the attenuation of the channel, the CM of the whole system $ABC$ becomes
\begin{align}
\gamma_3=\left[
\begin{matrix}
aI_2  & b\sqrt{\eta}\sigma_z & 2x\sigma_z  \\
b\sqrt{\eta}\sigma_z & (a\eta+(1-\eta)N_0)I_2 & -2x\sqrt{\eta}I_2 \\
2x\sigma_z  & -2x\sqrt{\eta}I_2 & (1+4x)I_2 \\
\end{matrix}
\right],
\end{align}
where $N_0$ is the variance of channel thermal noise.
In normal QKD protocols, Bob performs homodyne or heterodyne detection on the received signals. However, the direct-detection scheme may leave the attacker loophole to eavesdrop information.
Instead, Bob prepares a vacuum state and applies a displacement operation on it. Using a balanced beam splitter, Bob mixes the incoming mode with his own mode.
The second balanced beam splitter transforms the CM into $\gamma_4=B_{BC}\cdot\gamma_3\cdot B^T_{BC}$. After the beam splitter, one of
the outputs is detected with homodyne detection using the self-reference technique, while another one is discarded directly.
The CM of the system $AB$ is
\begin{align}
\gamma_{AB}=\left[
\begin{matrix}
aI_2  & \frac{2x+b\sqrt{\eta}}{\sqrt{2}}\sigma_z  \\
\frac{2x+b\sqrt{\eta}}{\sqrt{2}}\sigma_z & \frac{1+N_0+4x(1-\sqrt{\eta})+a\eta-N_0\eta}{2}  \\
\end{matrix}
\right],\label{14}
\end{align}
which can be used for calculating the secret key rate of the protocol.

\subsection*{B. Attacking strategy with general eavesdropping}

A QKD protocol is secure against general attack when it is secure against Gaussian collective attack~\cite{PhysRevLett.114.070501,Leverrier2017Security}.
This part performs an asymptotic security analysis based on infinitely-many uses of the channel under Gaussian collective attack. In each transmission, Eve may intercept the mode
and make it interact with an ensemble of ancillary vacuum modes via a general unitary operation. One of the output modes is sent to Bob
while another one is stored in Eve's quantum memory (QM). These states in QM will be measured at the end of the protocol collectively.
Taking reverse reconciliation into account, the final key rate can be derived as
\begin{align}
R=\xi I_{AB}-\chi_{BE},
\end{align}
where $\xi$ denotes the reconciliation efficiency.
We can compute the mutual information in terms of signal-to-noise ratio as
\begin{align}
I_{AB}=\log_2{\frac{\varphi+1}{\omega}}.
\end{align}
$\varphi$ is the modulation variance in shot-noise units and $\omega$ represents the equivalent noise.
In the previous CVQKD protocols, Eve's system $E$ purifies $AB$, so that $S(E)=S(AB)$, and $S(AB)$ can be calculated from the symplectic
eigenvalues of the covariance matrix $V_{AB}$. In order to calculate the Holevo bound between Alice and Bob with the simplification of the expression, we denote the CM of the reduced state of systems $AB$ as~\cite{holevo1973holevo}
\begin{align}
\gamma_{AB}=\left[
\begin{matrix}
aI_2  & c\sigma_z \\
c\sigma_z & bI_2  \\
\end{matrix}
\right].
\end{align}
The symplectic eigenvalues can be calculated as~\cite{serafini2004symplectic}
\begin{align}
\nu^2_{1,2}=\frac{1}{2}[\Delta\pm\sqrt{\Delta^2-4D^2}],
\end{align}
where $\Delta=a^2+b^2-2c^2$ and $D=ab-c^2$. Moreover, the symplectic eigenvalue of the conditional CM $V_{B|A}$ is $\nu^2_3=b(b-c^2/a)$.
Therefore, we have $S(AB)=G(\nu_1)+G(\nu_2)$ and $S(B|A)=G(\nu_3)$ with
\begin{align}
G(x)=\left(\frac{x+1}{2}\right)\log_2{\left(\frac{x+1}{2}\right)}-\left(\frac{x-1}{2}\right)\log_2{\left(\frac{x-1}{2}\right)}.
\end{align}
Consequently, the information eavesdropped by Eve can be bounded by $\chi_{BE}=S(AB)-S(B|A)$.

\subsection*{C. Secret key rate of the separable-state CVQKD}
It is necessary to note that the proposed protocol is different from the traditional protocol as the above-involved states are displaced before being mixed on the beam splitter.
Without the displacement, the output of the first beam splitter is equivalent to a two-mode squeezed vacuum state. Another difference from the entanglement-based scheme is
that Bob injects the received mode and his own mode into one beam splitter instead of performing homodyne or heterodyne detection directly.
As analyzed in Sec.~\ref{sec2}, all these efforts are to keep the ancillary mode separable from system $AB$ while completing the task of
distribution entanglement between Alice and Bob. Whereas, in the traditional CVQKD system, the information is encoded on the mode that is sent to the channel under Eve's control.
Eve may hide her attack in the channel noise. It has been assumed that Eve's system purifies $AB$, which implies that $S(E)=S(AB)$.

In the proposed protocol,
the auxiliary mode used for distributing information is separable from $AB$. Alice's and Bob's labs as well as the classical communication are out of Eve's touch.
Namely, Eve cannot steal any information by attacking the ancilla, leading to
$S_{E}=0$.
A problem about upper bound arises. In~\cite{Takeoka2014Fundamental,wilde2017converse,Pirandola2017Fundamental}, it has been proved
that the secret key rate cannot be unbounded with increasing signal energy for normal CVQKD protocol~\cite{PhysRevLett.88.057902}.
The secret key rate satisfying the condition
\begin{align}
R\leq I_{AB}-\chi_{BE}\leq G(\varphi)-G(\nu_1)-G(\nu_2).
\end{align}
The limit for $\varphi\rightarrow+\infty$ for the right part of the inequation is regular and finite~\cite{Takeoka2014Fundamental,wilde2017converse,Pirandola2017Fundamental}.
The secret key rate will not be unbounded with increasing signal energy even though $\chi_{BE}$ is removed.
A positive multiple of sum of the projectors is added to smear the entanglement between the $C$ and $AB$ before transmission.
The displacement which is proportional to the modulation variance also appears in the noise.
The secret key rate of this scheme will not be unbounded as the signal-to-noise ratio is bounded regardless of the increasing signal energy.
The advantage of keeping the ancillary state separable is the displacement before beam splitter. Bob uses a displaced state to interact with the ancilla rather than
detects it directly. This operation is just to cut off Eve's disturbance. Then the secret key rate can be expressed as
$
R=\xi I_{AB},
$
where $\xi$ is the negotiation efficiency and $I_{AB}$ can be calculated from the CM of system $AB$ in Eq.~(\ref{14}).

\section{Simulation results}
\label{sec4}
\begin{figure}[htp]
  \centering
  \includegraphics[height=55mm]{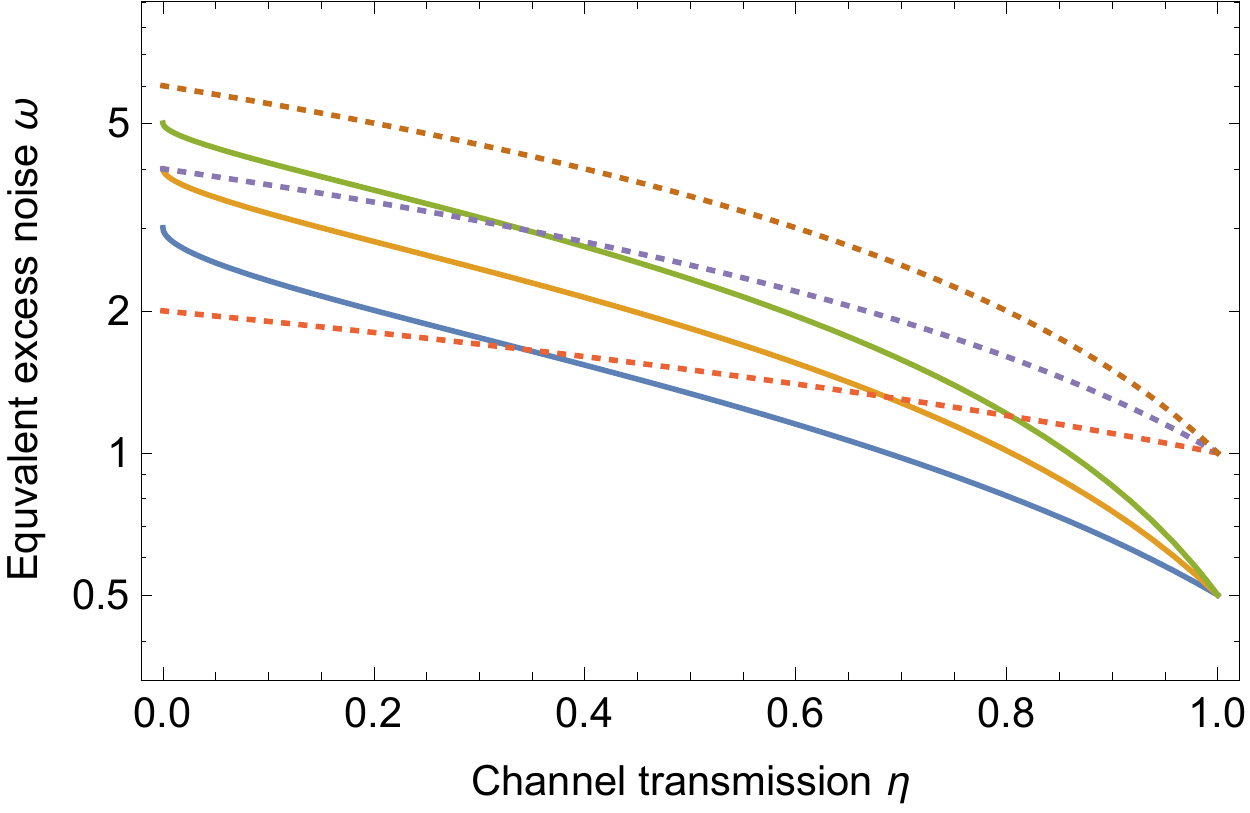}
  \caption{(Color online) Equivalent excess noise as a function of channel transmission $\eta$. The dashed lines are the equivalent excess noise of original protocol
  while the full lines denote the proposed one. From bottom to top, $N_0=1,3,5$.}
  \label{figure8}
\end{figure}
As discussed above, Alice and Bob can get the reduced CM $\gamma_{AB}$, from which they can calculate the secret key rate $R$ in Eq.(15).
Based on the Eq.~(\ref{14}), the equivalent excess noise can be expressed as
\begin{align}
\omega=\frac{1+(1-\eta)N_0+4x(2-\sqrt{\eta})}{2},
\end{align}
which is plotted in Fig.~\ref{figure8}. Compared with the traditional CVQKD protocol, the proposed protocol has an extra noise that is caused by the displacement operation.
The displacement may decrease the key rate $I_{AB}$. Fortunately, it can also remove the entanglement between the ancillary particle and the kept particles.

To demonstrate the performance of the protocol, we consider both direct reconciliation and reverse reconciliation.
\begin{figure}[!h]
  \centering
  \includegraphics[height=55mm]{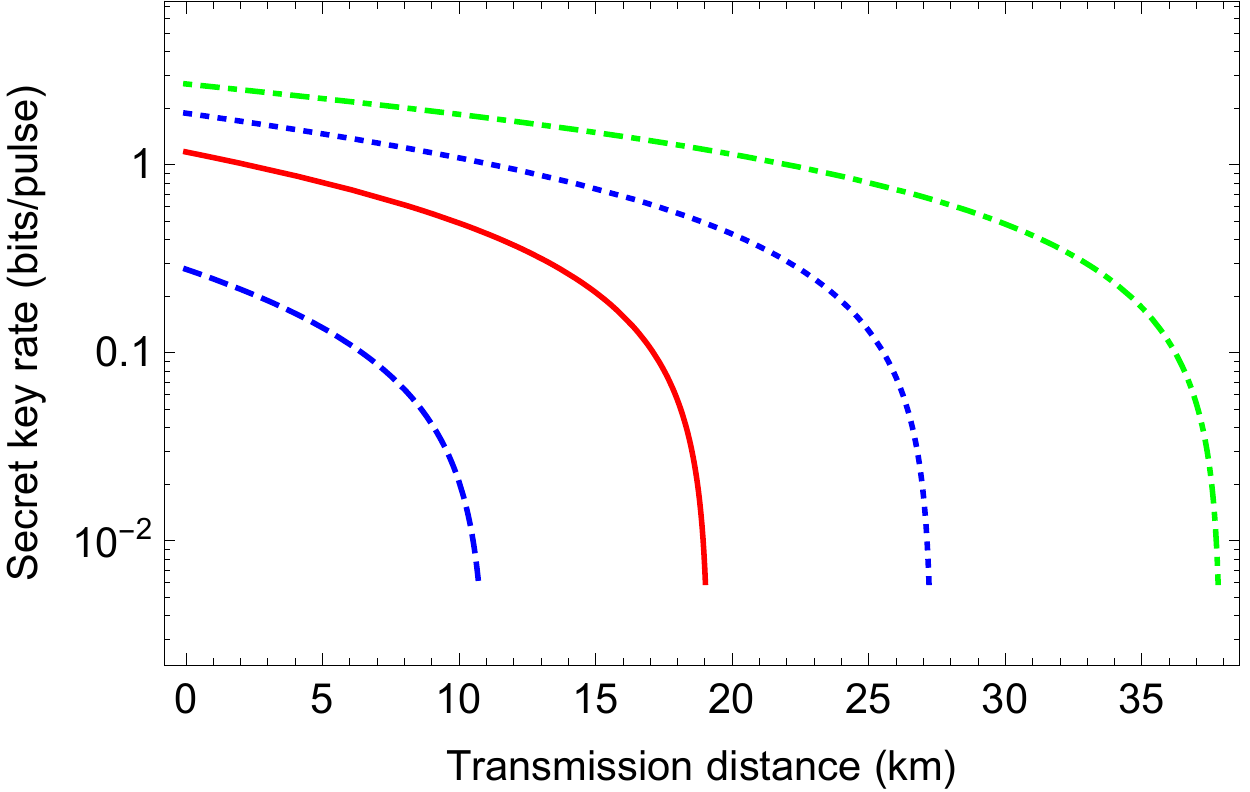}
  \caption{(Color online) Secret key rates versus transmission distance from Alice to Bob of the direct reconciliation case. The secret key rate decreases as the grow of the transmission distance.
  Simulation results refer to $V=2$ (blue dashed line), $V=10$ (red full line), $V=30$ (blue dashed line) and $V=100$ (green dot-dashed line).}
  \label{figure3}
\end{figure}
In Fig.~\ref{figure3}, we show the secret key rate of the proposed protocol with direct reconciliation. From top to bottom, the dashed, full, dotted and dot-dashed lines refer to the modulation variances
$2, 10, 30$ and $100$, respectively. With current technology, the $15\mathrm{dB}$ squeezed states of light has already been detected in~\cite{PhysRevLett.117.110801}.
The transmission can exceed $15\mathrm{km}$, which corresponds to the $3\mathrm{dB}$ restriction in direct reconciliation.
Moreover, the excess noise has been taken into consideration with $\epsilon=0.05$ and reconciliation efficiency is set $\beta=0.95$ for all numerical simulations.

\begin{figure}[!h]
  \centering
  \includegraphics[height=55mm]{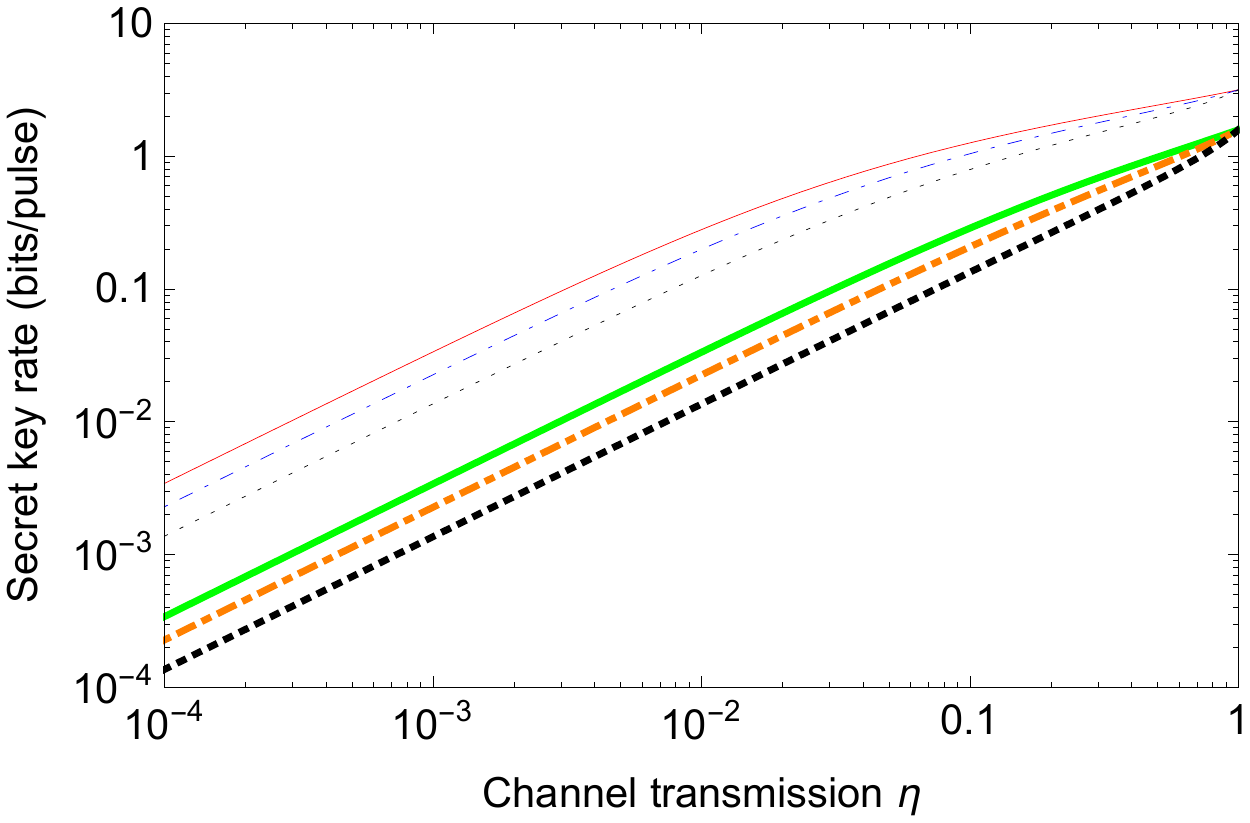}
  \caption{(Color online) Secret key rates versus channel transmission, $\eta$. The full lines are under the ideal condition with zero excess noise
  while the dot-dashed and dotted lines correspond to $N_0=2$ and $4$, respectively. The thick and thin lines are under the condition that modulation
  variance $V=10$ and $100$.}
  \label{figure4}
\end{figure}
The simulation result in Fig.~\ref{figure4} is the secret key rate of the direct reconciliation case. The difference between thin lines and thick lines shows
that modulation variance plays a positive role in the secret key rate. However, the displacement term limits the continued increase of the secret key rate.
The full line, dot-dashed line and dotted line show channel noise has a negative effect on the secret key rate.
We find that there is little effect of the noise on the secret key rate of the CVQKD system when the transmittance approaches to one.

\begin{figure}[!h]
  \centering
  \includegraphics[height=55mm]{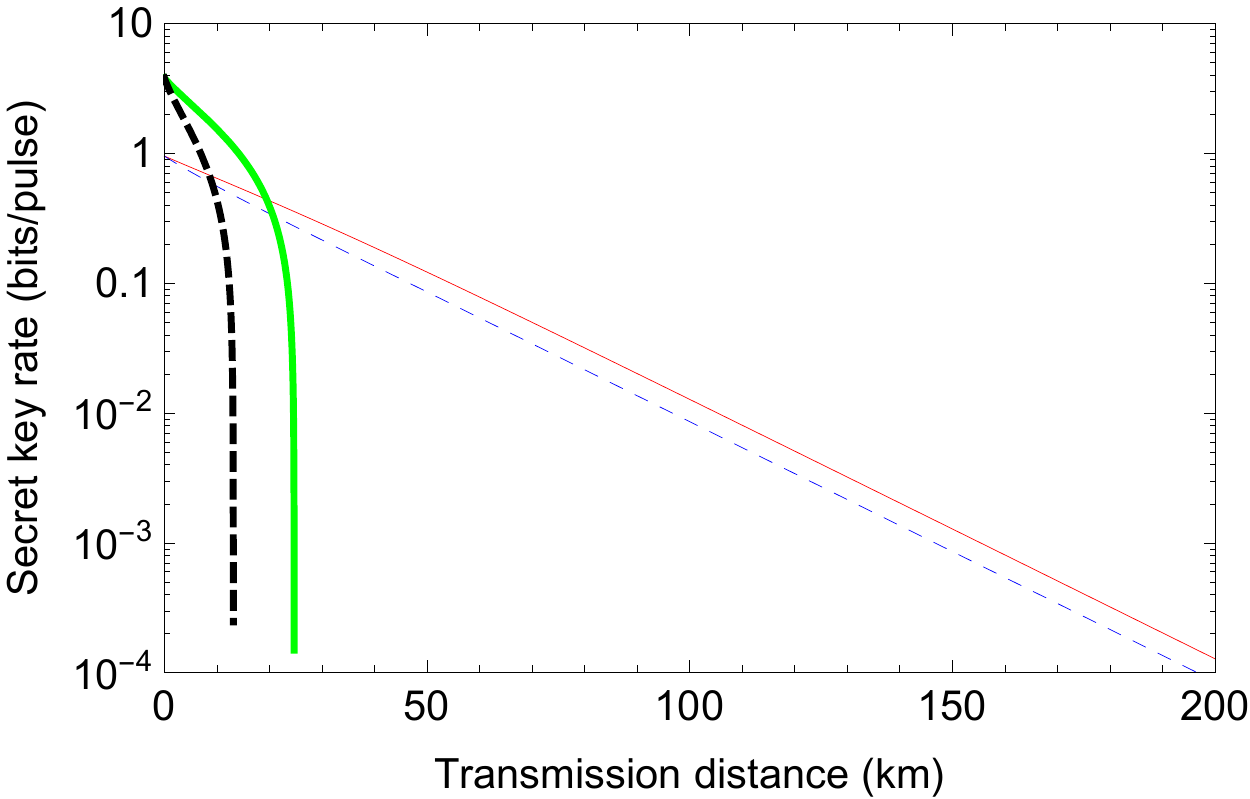}
  \caption{(Color online) Secret key rates versus transmission distance, $L$. The full lines correspond to the condition with excess noise $N_0=1.01$
  while the dashed lines correspond to $N_0=2$. The thin lines represent the proposed protocol with separable Gaussian states while the thick lines are the
  the traditional protocols. In the simulation, the modulation variance $V=30$.}
  \label{figure5}
\end{figure}

Fig.~\ref{figure5} demonstrates the secret key rates of the proposed protocol using a separable ancilla in the reverse reconciliation case.  The traditional CVQKD system can only transmit $30\mathrm{km}$ due to the existence of the eavesdropper, whereas the proposed protocol achieves the transmission distance $200\mathrm{km}$ at rate of $10^{-4}$ bits per pulse. The transmission distance of the separable-state CVQKD protocol is lower than that of the traditional one. This phenomenon may result from the abandon of the ancillary particle.
Moreover, we can also find that the protocol has a better tolerance to noise than the traditional one.

\begin{figure}[!h]
  \centering
  \includegraphics[height=55mm]{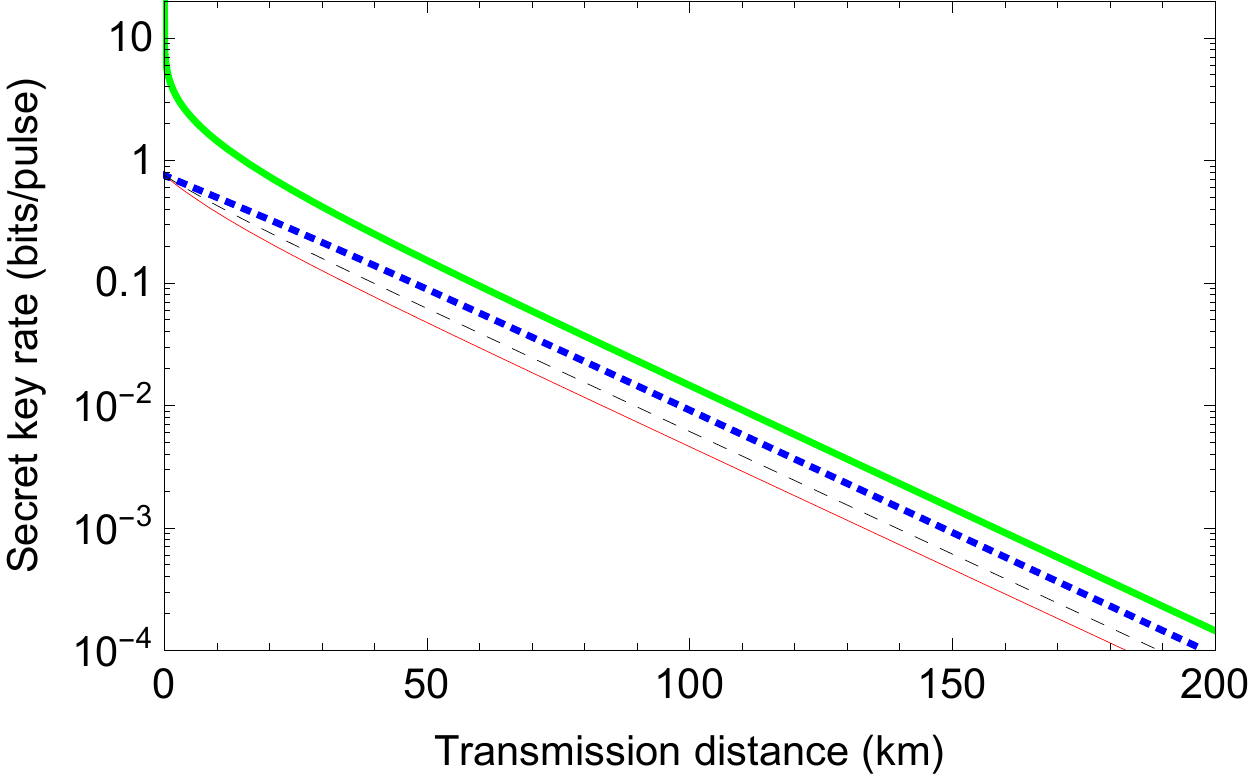}
  \caption{(Color online) Secret key rates of CVQKD with separable states versus the upper bound of CVQKD. The thick green line is the upper bound of the traditional CVQKD. The dotted, dashed and thin full lines
  are the proposed CVQKD protocols with $N_0=1,2,3$, respectively.}
  \label{figure9}
\end{figure}
In Fig.~\ref{figure9}, we make a comparison between the secret key rate of our protocol and the fundamental limit~\cite{Pirandola2017Fundamental,PhysRevLett.102.050503}.
~\cite{Pirandola2017Fundamental} proved the PLOB bound, while ~\cite{wilde2017converse} later discussed the strong convergence of this bound.
The top green line is the fundamental limit of general CVQKD protocol, which is given by $-\log_2(1-\eta)$. $\eta$ is channel transmittance of the pure-loss channel.
As shown in~\cite{wilde2017converse,Pirandola2017Fundamental}, the protocols whose secret key rate is based on the lower bound cannot come up with the upper bound
when the transmittance $\eta$ is less than $0.7$.
The protocol based on transmission of separable Gaussian states via a quantum channel and LOCC operation has a good performance on the aspect of transmission distance.
This scheme has a good tolerance for excess noise and the transmission distance achieves $200\mathrm{km}$.

\section{Conclusion}
\label{sec5}
We have proposed an improved continuous-variable quantum key distribution protocol that is immune to Eve's attack.
This separable-state CVQKD protocol is different from the traditional protocol because the ancillary particle is separable from Alice and Bob¡¯s system.
In previous protocols, the information is encoded on the particles which will pass through a quantum channel controlled by Eve.
Eve can purify the whole system and extracted as much information as the Holevo bound of the system.
In addition, after the two respective particles interact continuously with an ancilla, they get entangled, leaving the ancilla separable all the time.
The displacement operation in the preparation course plays a crucial role in smearing the
entanglement between the ancilla and Alice and Bob's system. The secret key rate of the separable-state CVQKD will not be
unbounded with increasing signal energy. The proposed protocol has good tolerance to extra noise and is
able to keep abreast of the upper bound until $200\mathrm{km}$.
We note that the proposed CVQKD protocol can be practically implemented using separable Gaussian states as entanglement
preparation processes based on separable Gaussian states have been demonstrated in experiment~\cite{PhysRevLett.111.230506,PhysRevLett.111.230504}.

\section*{Acknowledgements}

We would like to thank L. Mi$\check{\mathrm{s}}$ta for helpful discussion. This work is supported by the National Natural Science Foundation of China (Grant Nos. 61572529) and the Fundamental
Research Funds for the Central Universities of Central South University (2017zzts144).

\bibliographystyle{apsrev4-1}
\bibliography{reference}
\end{document}